\documentclass[prb,aps,twocolumn,superscriptaddress,10pt,showpacs,longbibliography,noeprint]{revtex4-1}
\usepackage{graphicx}
\usepackage{epstopdf}
\usepackage{color}
\usepackage{amsmath,amssymb}
\usepackage{fixmath}

\newcommand{\angstrom}{\mbox{\normalfont\AA}}
\usepackage [english]{babel}
\usepackage [autostyle, english = american]{csquotes}
\MakeOuterQuote{"}
\usepackage{textgreek}
\usepackage[export]{adjustbox}
\usepackage{siunitx}
\usepackage{xcolor}
\usepackage{hyperref}
\usepackage{notes2bib}
\begin{document}
\title{Soft mode anisotropy in negative thermal expansion material ReO$_3$}	
\author{Tobias A. Bird} 
\affiliation{Department of Chemistry, University of Warwick, Gibbet Hill, Coventry, CV4 7AL,United Kingdom}
\author{Mark G. L. Wilkinson}
\affiliation{School of Physics and Astronomy, Queen Mary University of London, London, E1 4NS, United Kingdom}

\author{David A. Keen}  
\affiliation{ISIS, Rutherford Appleton Laboratory, Harwell Campus, Didcot, Oxfordshire OX11 0QX, United Kingdom}
\author{Ronald I. Smith}
\affiliation{ISIS, Rutherford Appleton Laboratory, Harwell Campus, Didcot, Oxfordshire OX11 0QX, United Kingdom}

\author{Nicholas C. Bristowe}
\affiliation{Centre for Materials Physics, Durham University, South Road, Durham DH1 3LE, United Kingdom}

\author{Martin T. Dove}
\affiliation{Schools of Computer Science and Physical Science \& Technology, Sichuan University, Chengdu 610065, People's Republic of China}
\affiliation{Department of Physics, School of Sciences, Wuhan University of Technology, 205 Luoshi Road, Homgshan district, Wuhan, Hubei 430070, People's Republic of China}
\affiliation{School of Physics and Astronomy, Queen Mary University of London, London, E1 4NS, United Kingdom}

\author{Anthony E. Phillips}
\email{a.e.phillips@qmul.ac.uk}
\affiliation{School of Physics and Astronomy, Queen Mary University of London, London, E1 4NS, United Kingdom}

\author{Mark S. Senn}
\email{m.senn@warwick.ac.uk}
\affiliation{Department of Chemistry, University of Warwick, Gibbet Hill, Coventry, CV4 7AL,United Kingdom}

\date{\today}

\begin{abstract}
We use a symmetry-motivated approach to analyse neutron pair distribution function data to investigate the mechanism of negative thermal expansion (NTE) in ReO$_3$. This analysis shows that the local structure of ReO$_3$ is dominated by an in-phase octahedral tilting mode and that the octahedral units are far less flexible to scissoring type deformations than the octahedra in the related compound ScF$_3$. These results support the idea that structural flexibility is an important factor in NTE materials, allowing the phonon modes that drive a volume contraction of the lattice to occupy a greater volume in reciprocal space. The lack of flexibility in ReO$_3$ restricts the NTE-driving phonons to a smaller region of reciprocal space, limiting the magnitude and temperature range of NTE. In addition, we investigate the thermal expansion properties of the material at high temperature and do not find the reported second NTE region. Finally, we show that the local fluctuations, even at elevated temperatures, respect the symmetry and order parameter direction of the observed $P4/mbm$ high pressure phase of ReO$_3$. The result indicates that the motions associated with rigid unit modes are highly anisotropic in these systems. 

\end{abstract}

\maketitle	

\section{Introduction}
The phenomenon of negative thermal expansion (NTE) is an intriguing and unusual property for a material to exhibit. Broadly speaking, there are two families of NTE materials: those in which the anomalous thermal expansion behaviour arises solely from vibrational effects, and those in which it has an electronic origin\cite{Attfield2018,Lind2012,Barrera2005,Dove2016}. In a typical material, we expect to observe positive thermal expansion (PTE) since the anharmonic shape of the interatomic potential leads to an increase in the equilibrium distance between two atoms as they gain more energy from an increase in temperature. The typical explanation for how vibrational effects can lead to a deviation from this behaviour is the "tension effect"\cite{Attfield2018,Dove2016}. A lot of structural NTE materials consist of a network of cation-anion linkages, often with an anion linked to two cations in a straight line\cite{Kennedy2005,Tiano2003,Fang2013,Gao2018}. If the energy cost to expand these bonds is quite high, then the central ion can displace perpendicularly to the bonds, which has the effect of pulling the two outer ions towards each other. The bonds connecting the linked ions still show PTE, but the linkage as a whole shrinks. This hypothesis is supported by neutron and X-ray powder diffaction experiments showing significant transverse atomic displacement parameters\cite{Dove2016}. The tension effect is often realised in NTE materials \textit{via} rigid unit modes (RUMs)\cite{Dove2016,Dove1999} -- materials made from a network of rigid polyhedra often have a high energy cost to distort the polyhedral units but a low energy cost for cooperative rotations of the polyhedra. These distortions typically cause a contraction of the volume of the material and since they are low in energy will have a large contribution to the overall thermal expansion behaviour. Many of these cation-anion linked NTE materials give rise to a network of connected polyhedra, such as the archetypal NTE material ZrW$_{2}$O$_{8}$\cite{Evans1996}, SiO$_2$\cite{Dove1998} and layered materials showing uniaxial NTE such as Ca$_3$Mn$_2$O$_7$\cite{Senn2015}.

Two materials that are often used to illustrate the RUM model are ReO$_3$ and ScF$_3$, due to their relatively simple structure when compared to more complex materials like ZrW$_{2}$O$_{8}$. Both compounds consist of corner-sharing octahedra and both exhibit NTE, up to around 220 K in ReO$_3$\cite{Chatterji2008,Chatterji2009} and 1100 K in ScF$_3$\cite{Greve2010}, although the exact range of NTE in the former is dependent on sample preparation\cite{Rodriguez2009}. There has also been an observation of a reappearance of NTE in ReO$_3$ between 600--700 K\cite{Chatterji2009}. This overlaps with the temperature at which ReO$_3$ is known to decompose to Re$_2$O$_7$ and ReO$_2$\cite{Schulman1938}. Both ReO$_3$ and ScF$_3$ undergo a phase transition \textit{via} octahedral tilts with applied pressure. In ReO$_3$, the octahedral tilts in successive planes along the tilt axis are in-phase tilts\cite{Axe1985,Jorgensen1986,Chatterji2006,Liu2015}, whereas for ScF$_3$, the tilts are out of phase\cite{Greve2010,Aleksandrov2009}.

Compared to typical ABX$_3$ perovskites, which also exhibit rigid rotations of the octahedral units, ReO$_3$ and ScF$_3$ both have a vacant $A$-site, allowing larger tilt angles\cite{Jorgensen1986}. Recent analysis of the local structure of ScF$_3$\cite{Bird2020,Dove2020} demonstrated that flexibility of the octahedra themselves, \textit{i.e.}, distortions of the intraoctrahedral F--Sc--F bond angles away from 90$^{\circ}$, is a key contributor to the NTE. In perovskites, pure RUMs are restricted to the line M-R in reciprocal space, which in principle gives their mode Gr{\"u}neisen parameters a vanishingly small contribution towards the mean Gr{\"u}neisen parameter. On moving away from this line, the modes have an increasing component of octahedral distortion. These modes, termed quasi-RUMs, can still contribute to NTE if they have negative Gr{\"u}neisen parameters. Simplistically, structures with greater flexibility will have a greater volume of reciprocal space occupied by quasi-RUMs with a negative Gr{\"u}neisen parameter than those of lesser flexibility. In addition, these modes will have a greater contribution to the mean Gr{\"u}neisen parameter, since their component of octahedral deformation will have a lower energy cost\cite{Dove2019a}. The oxygen anions in ReO$_3$ have an increased charge compared to the fluoride anions in ScF$_3$ and therefore an increased coulomb repulsion force between them, which one might expect to decrease the flexibility of the octahedra. Molecular dynamics simulations of an ReO$_3$-like structural model have also shown a decreasing magnitude of NTE for increasing anion interaction strengths\cite{Schick2016}. In particular, the sign of the coefficient of thermal expansion in ScF$_3$ has been shown to be highly sensitive to changes in the force constant governing flexing of the F$-$Sc$-$F right angle\cite{Dove2020}.

We investigate this hypothesis by using a symmetry-motivated approach to analyze neutron pair distribution function (PDF) data collected on ReO$_3$ across its entire temperature range of stability. Our study enables us to identify the characters of the dominant dynamic deviations away from the average structure as a function of temperature.

\begin{figure}[h!]
	\centering
	\includegraphics[width=\columnwidth]{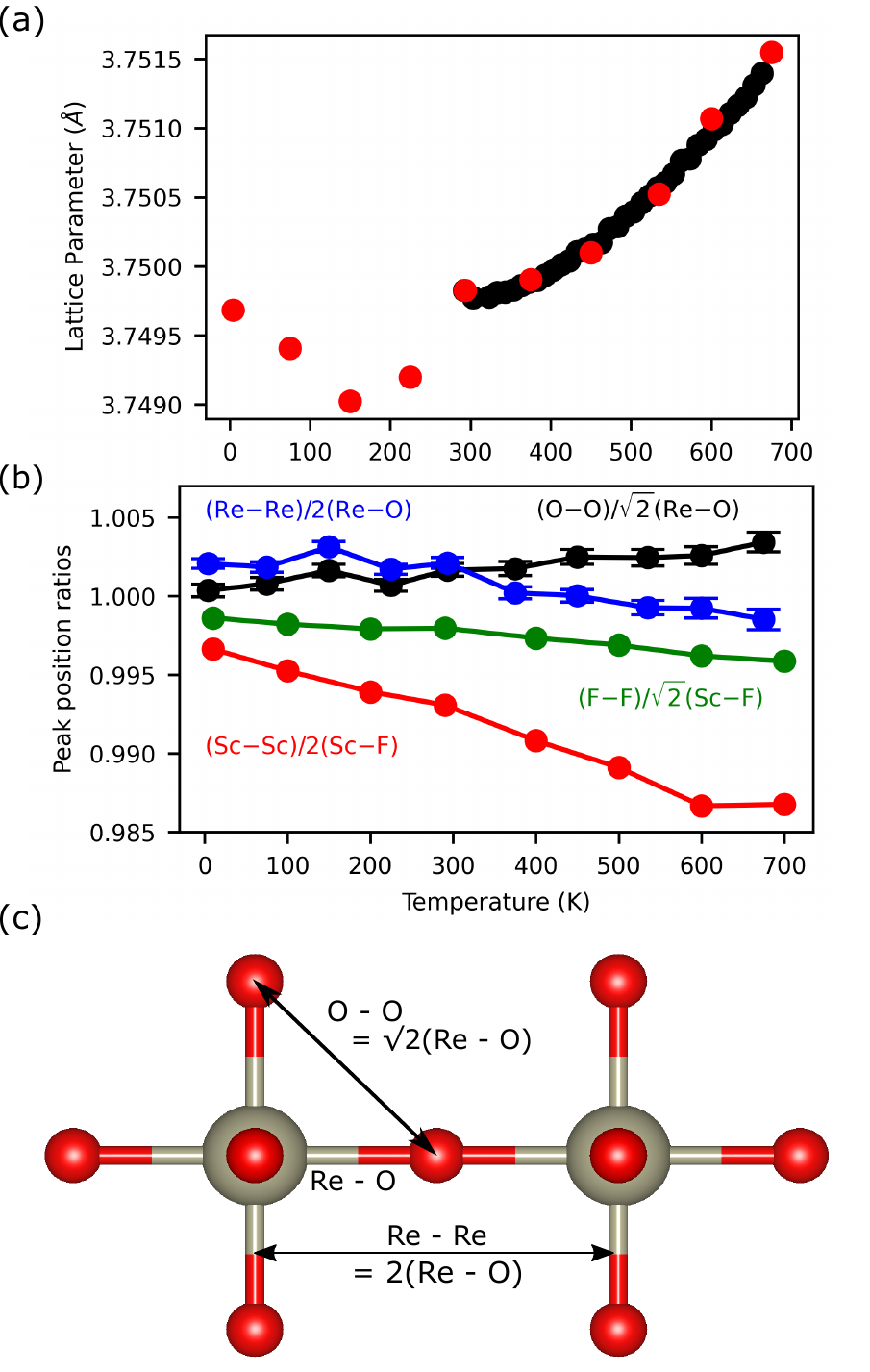}
	\caption{\label{presapa}(a) Temperature variation in the cubic lattice parameter of ReO$_3$ from Rietveld refinement. The data points shown in red are from the longer collections which were used for total scattering measurements. (b) Plots of the PDF peak position ratios indicated in the legend. Peak positions for ScF$_3$ were extracted from atomistic configurations generated using the reverse Monte Carlo method on neutron PDF data\cite{Dove2020} and OriginLab was used to fit the peaks for ReO$_3$. The error bars for ReO$_3$ correspond to the uncertainties from the peak fitting algorithm. We note that values of (Re--Re)/2(Re--O) above 1 are not physically possible, but at worst they are less than 0.05 \% too high. This is likely due to a systematic error and we believe that the trends with temperature are reliable. The uncertainty in these atom pair distances due to the r-space resolution ($\Delta r \approx \SI{0.1}{\angstrom}$) is an order of magnitude greater than this. (c) A diagram showing how the O--O and Re--O distances in the average structure of ReO$_3$ are related to one another.}
\end{figure}

\section{Experimental Details}

Rhenium trioxide was purchased from Sigma-Aldrich and used as received. A 4.63965 g sample was loaded into a vanadium can of 6 mm diameter and mounted onto the Polaris instrument at ISIS Neutron and Muon Source (Rutherford Appleton Laboratory, U.K.)\cite{Smith2019}. Data were collected in “short” 13 $\mu$A hr runs for Rietveld refinement and “long” sets of five 150 $\mu$A hr runs for pair distribution function analysis, equivalent to data collection periods of 5 minutes and 1 hour for each run, respectively. For low temperatures (between 4 K and 293 K) a helium flow cryostat was used, whereas for higher temperatures from 293 K to 750 K a furnace was used. The furnace data were collected in two separate experiments, due to an unscheduled beam shut-off at the facility during the first experiment. The furnace data for 600 K and above were taken from the second experiment. Both experiments used the same ReO$_3$ sample, stil loaded in the same sample can. During these experiments, data were also collected from the empty instrument, empty sample environment, and an empty vanadium can for background correction, as well as a solid vanadium rod for normalisation. 
\begin{figure*}[t]
	\centering
	\includegraphics[]{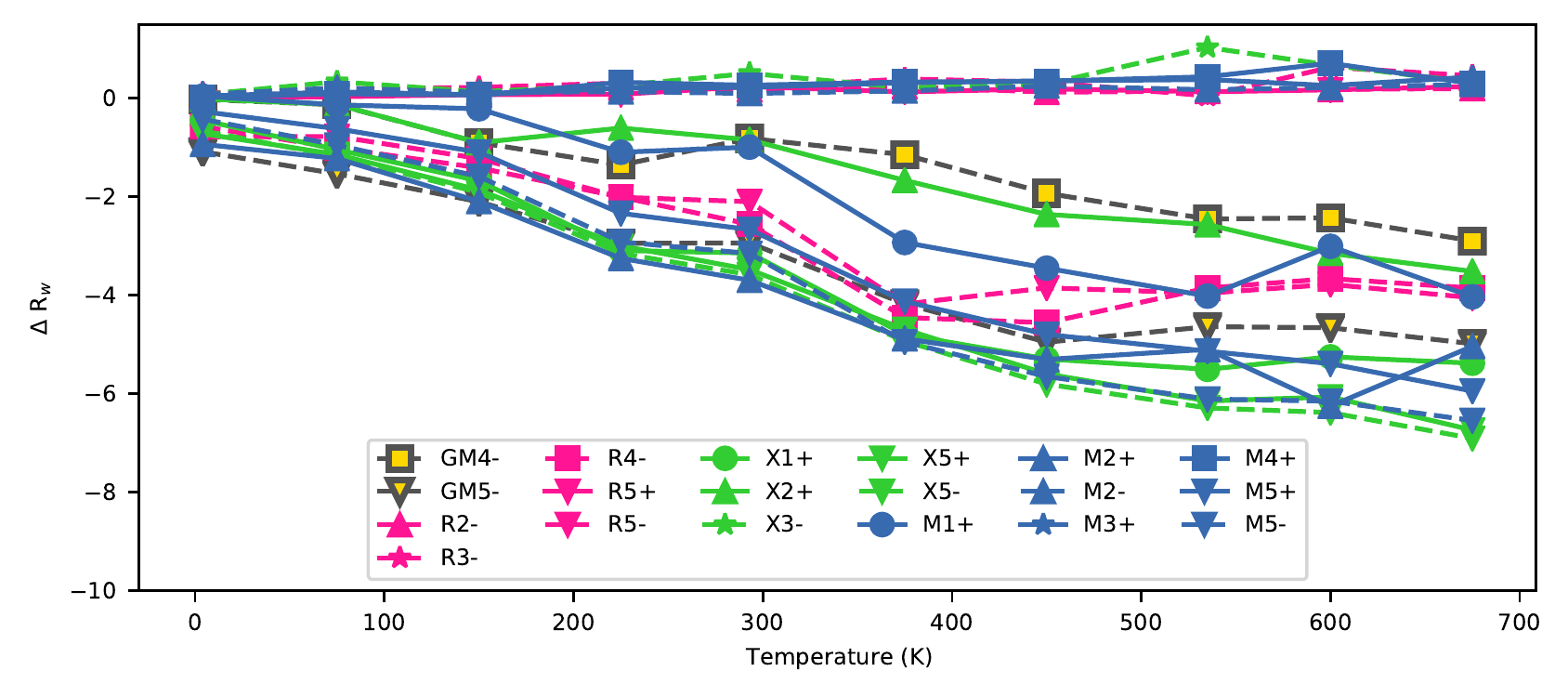}
	\caption{\label{sapa}The results of the SAPA on ReO$_3$. The best individual fitting statistic is plotted for each irrep at each temperature. The $R_{w}$ is shown relative to the $R_{w}$ for the refinement with no symmetry adapted displacement modes active. The irreps are labelled as follows: colour denotes the k-point of the irrep, with blue referring to the M point, green to X, pink to R and yellow to \textGamma; marker shape denotes the irrep number, with a circle referring to 1, an upward-pointed triangle to 2, a star to 3, a square to 4 and a downward-pointed triangle to 5; linestyle denotes the parity of the irrep, with a solid line referring to a $+$ irrep, and a dashed line referring to a $-$ irrep. A representative fit of the average structure to the PDF at 293 K is shown in the SI.}
\end{figure*}
\begin{figure}[]
	\centering
	\includegraphics[]{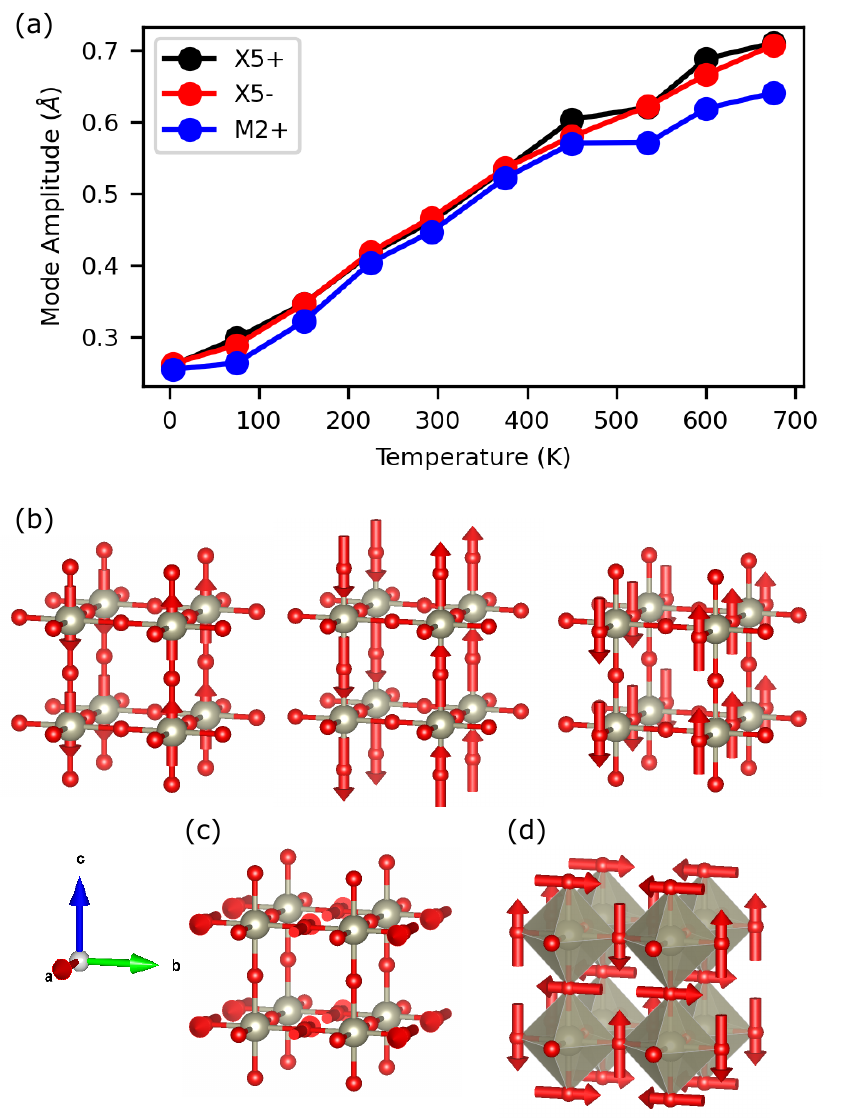}
	\caption{\label{modeamps}(a) Temperature variation of the normalised mode amplitudes of the modes belonging to the M$_2^+$, X$_5^+$ and X$_5^-$ irreps. The values were calculated from the lowest $R_{w}$ refinement for each temperature and irrep from the output of the SAPA. These values are supercell normalised mode amplitudes (A$_s$ as defined in ISODISTORT). To convert to the A$_p$ values, divide by a factor of $2\sqrt{2}$. Visualisations of the modes which transform as each irrep are shown below. Specifically, (b) the three distortions for the X$_5^+$ (a,0;0,0;0,0), (c) X$_5^-$ (a,0;0,0;0,0) and (d) M$_2^+$ (a;0;0).}
\end{figure}
For average structure determination, data reduction was carried out using Mantid\cite{Arnold2014}. The lattice parameter for each temperature was determined by Rietveld refinement using the EXPGUI\cite{Toby2001} interface to GSAS\cite{Larson2000} against the data from Polaris detector banks 3--5, refining both the unit cell and atomic displacement parameters (noting that all atomic coordinates for the Re and O atoms are fixed by symmetry in the $Pm\bar{3}m$ space group). A representative fit to the data at 150 K is included in the SI. The background was modelled using an 8-term shifted Chebyschev function. An absorption correction was refined to account for the neutron absorption of rhenium. 

To ensure that the unit cell parameters were consistent between the cryostat and furnace environments, data were collected at 293 K in both environments. For the furnace data set, the unit cell parameter was fixed at the value determined from the cryostat data and the diffractometer constant DIFC for the backscattering detector bank was refined instead. This new value was held constant for subsequent furnace data sets to ensure self-consistency across the entire temperature range (with an identical approach used to ensure consistency between the first and second furnace data sets). 

For local structure analysis, Gudrun\cite{McLain2012} was used to subtract the background, correct the data for self-shielding, absorption, and multiple scattering, normalise them to give the scattering function $S(Q)$, and finally Fourier transform this to give the pair distribution function $D(r)$. The scattering function $S(Q)$ was determined over the $Q$ range 0.6--50 \si{\angstrom}$^{-1}$ in steps of 0.02 \si{\angstrom}$^{-1}$. This data normalisation is also dependent on the density of the powdered sample: the powder packing fraction was initially set to the value measured experimentally and then adjusted by hand for all data sets in order to set the limiting value of the total scattering structure factor, $F(Q)$\bibnote{The total scattering structure factor, $F(Q)$, is related to the scattering function $S(Q)$ by the relation $S(Q) - 1 = F(Q)/\left(\sum_{i=1}^{n} c_i \bar{b}_i\right)^2$.} , as $Q \rightarrow 0$ to its theoretical value of $-\sum_{i} c_i \bar{b}_i^2$\cite{Keen2001}, and to set the coordination number for the first peak (Re--O) to its ideal value of six O atoms per Re atom.

Analysis of the pair distribution functions was carried out using the symmetry-adapted PDF analysis (SAPA) method described in ref. 23. For each sample, a 2 $\times$ 2 $\times$ 2 $P1$ supercell of the $Pm\bar{3}m$ aristotype ReO$_3$ with Re at (0.5, 0.5, 0.5) and O at (0.5, 0.5, 0) was generated and parameterised in terms of symmetry adapted displacements using the ISODISTORT software\cite{Campbell2006}. The modes modelled using this supercell expansion only represent a small fraction of possible phonon modes. However, even if the exact wave vectors of the soft phonon modes do not coincide with the \textGamma, X, M or R points of reciprocal space, since their characters should vary continuously between high symmetry points, the nature of the local symmetry breaking should still be manifested in our analysis. The generated mode listings were output in .cif format and then converted to the .inp format of the TOPAS Academic software v6 using the Jedit macros\cite{Evans2010}. In total, there were 96 modes which transformed according to one of 19 irreducible representations. For each irreducible representation (irrep) at each temperature, refinements of the corresponding modes were started from random starting mode amplitudes. This was repeated 500 times for each irrep at each temperature to ensure the global minimum of the refinement was reached. For all temperatures, the refinements were carried out with a fitting range of 1.5 to $\SI{10}{\angstrom}$. Refinements were carried out using the Topas Academic software v6\cite{Coelho2015}.

The DFT calculations were performed using the Vienna Ab Initio Simulation Package (VASP)\cite{Kresse1994,Kresse1996,Kresse1996b,Kresse1993}, version 5.4.4.
We employed the PBEsol exchange correlation potential\cite{Perdew2008} and projector augmented-wave (PAW) pseudopotentials\cite{Blochl1994,Kresse1996}, as supplied within the VASP package.
A plane wave basis set with a 900 eV energy cutoff and a $12 \times 12 \times 12$ Monkhorst-Pack k-point mesh with respect to the parent cubic primitive cell (scaled accordingly for other supercells) were found suitable.

\section{Results and Discussion}

Rietveld refinement of the $\mathbf{a}$ unit cell parameter from the Bragg reflection positions observed during the total scattering experiment shows we observe the same low temperature negative thermal expansion range as literature reports for ReO$_3$\cite{Chatterji2008,Chatterji2009} (Fig. \ref{presapa} (a)). The high temperature measurements for both long and short runs do not show the second region of NTE observed by Chatterji \textit{et al}\cite{Chatterji2009}. Our conjecture is that the original observation of this phenomenon was likely due to sample decomposition, since above 673 K, ReO$_3$ starts to decompose \textit{via} disproportionation\cite{Schulman1938}.

We can gain some insights into the local structure from the PDFs of ReO$_3$ without performing any modelling. In Fig \ref{presapa} (b), we quantitatively estimate the flexing of the O--Re--O right angle and Re--O--Re linkage. In the average structure, the plotted ratios are constrained to be constant at a value of one, due to a 90$^{\circ}$ O--Re--O bond angle and a straight Re--O--Re bond. The intercepts of these ratios at 0 K reflect the inherent flexibility of the structures, since the force constants governing both types of flexing motion do not change with temperature. This shows that ScF$_3$ is more flexible than ReO$_3$, presumably due to the lower charge of F$^-$ compared to O$^{2-}$. The difference in flexibility with respect to distortions of the O--Re--O (F--Sc--F) bond angle is smaller than that for distortions of the Re--O--Re (Sc--F--Sc) linkage. However, molecular dynamics simulations have shown that the sign of the coefficient of thermal expansion is more sensitive to changes in the force constant governing the former\cite{Dove2020}. Another difference of note is that the trends in the $X$--$M$--$X$ bond angle for ReO$_3$ and ScF$_3$ are opposed. The average O--Re--O bond angle increasing with temperature while the F--Sc--F bond angle decreases. These differing trends could reflect that phonon modes with differing characters and amplitude cause the octahedral deformations.

To gain a fuller understanding of the character of the soft modes in ReO$_3$, we turn to the results from our SAPA method. In Fig \ref{sapa}, we can see there is a group of irreducible representations whose modes consistently give the best improvement in $R_{w}$ compared to the average structure refinements at each temperature. These are M$_2^+$, M$_5^+$, M$_5^-$, X$_5^+$, X$_5^-$ and \textGamma$_5^-$. All but M$_2^+$ have symmetry-adapted displacements associated with them which are of a scissoring character: the Re-O octahedral bond lengths remain unchanged, but some of the bond angles are distorted \textit{via} transverse displacements of the O anions. The M$_5^-$ and X$_5^+$ irreps both also support bond-stretching. However, this type of distortion is relatively high in energy so contributes negligibly both to the overall coefficient of thermal expansion and to the refined distortion. The amplitudes of these scissoring modes (Fig \ref{modeamps} (a)) are smaller than for ScF$_3$\cite{Bird2020} which further supports the hypothesis that a lower flexibility is responsible for the reduced magnitude of NTE in ReO$_3$ compared to ScF$_3$.

\begin{figure}[t]
	\centering
	\includegraphics[]{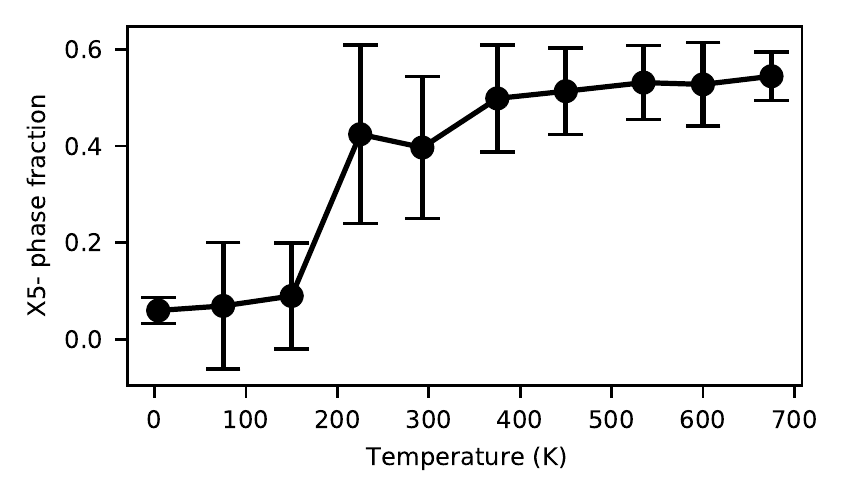}
	\caption{\label{m2ppc} A plot of the weighted mean phase fraction of the X$_5^-$ phase from two-phase refinements of M$_2^+$ in one phase and X$_5^-$ OPD in the other. Similarly to the SAPA, 500 repeats were performed at each temperature from randomised starting values and the weighting, $w$, of each refinement towards the mean was calculated according to $w = \exp[(R_{min}-R_i)/0.1]$, where $R_{min}$ is the minimum $R_{w}$ for each temperature and the $R_i$ are the $R_{w}$ of each refinement.}
\end{figure}

The other local symmetry breaking which is consistently amongst those which show the most improvement in $R_{w}$ from the average structure, and at some temperatures shows the most improvement overall, transform as irrep M$_2^+$. The distortions belonging to this irrep correspond to an in-phase tilt of the octahedra, with no other type of distortion associated with it, so it is of pure RUM character. This is a significant distortion for ReO$_3$ which undergoes a phase transition with applied pressure (\textit{ca.} 5 kbar at 300 K) during which the M$_2^+$ mode softens and the tilts are "frozen in" to the structure. To compare the relevant prevalence of this pure RUM mode with the scissoring modes, we refined a two-phase model, with displacements from the X$_5^-$ irrep in one phase and from the M$_2^+$ in another. The X$_5^-$ irrep was chosen here since all the displacements associated with it are of a scissoring mode character and it provides a better fitting statistic than other irreps for which this is also true (M$_5^+$ and \textGamma$_5^-$, Fig \ref{sapa}). At temperatures below 225 K, in the negative thermal expansion region, the RUM dominates as evidenced by the refined phase fraction in Fig \ref{m2ppc} (a). At higher temperatures, there is a balance between the two phase fractions, indicating that both scissoring and RUM type motions account for a substantial proportion of the dynamic distortions. 

At first glance, this increase in the proportion of motion arising from scissoring modes, coincident with the onset of PTE, is perhaps contradictory to the quasi-RUM mechanism for NTE which has been previously discussed for ScF$_3$. In ScF$_3$, the octahedral flexibility means a significant proportion of quasi-RUMs, modes which are of mixed scissoring and rigid unit mode character, will have a negative Gr{\"u}neisen parameter. This expands the volume occupied in reciprocal space by NTE phonons, increasing the overall contribution of modes with negative mode Gr{\"u}neisen parameters to the mean Gr{\"u}neisen parameter. Previous analysis of both X-ray and neutron PDF data of ScF$_3$ finds that the majority of motion of fluorine atoms comes from scissoring modes, with approximate scissoring:RUM ratios of 4:1 (0.8 as a phase fraction) from SAPA analysis\cite{Bird2020} and 3.5:1 (0.78) from reverse Monte Carlo analysis\cite{Dove2020} at all temperatures. Our analysis in Fig \ref{m2ppc} shows that in ReO$_3$, the scissoring:rotation ratio is close to 1:1 at 675 K, but drops to 1:10 by 4 K. These ratios suggest that there is significantly more resistance to octahedral deformations in ReO$_3$ than ScF$_3$. However, within renormalized phonon theory, we would expect the phase fractions\bibnote{It is important to note here that the phase fractions we discuss simply reflect an approximate measure of the proportion of the motion of O anions arising from the two different distortions. We do not believe that the two distortions exist in distinct phases within the material.} in Fig \ref{m2ppc} to remain constant with temperature, as we observe in our RMC analysis of ScF$_3$. The fact that the proportion of the scissoring to RUM phase fraction increases at high temperatures is more likely to reflect a hardening of the RUM than a sudden increase in the flexibility of the structure. This supposition is supported by the observation that the pressure at which the first phase transition occurs, increases with temperature\cite{Chatterji2006}. The increased energy of the phonon modes with RUM character on warming lowers their contribution to the mean Gr{\"u}neisen parameter at a given temperature. This hardening also has a knock-on effect on the quasi-RUMs, since their RUM component will also have a higher energy. It is of course tempting to suggest that this hardening of the modes with RUM character above 200 K is responsible for the concurrent change in sign of the bulk thermal expansion coefficient. However, it is conversely evident from the negative sign of the Gr{\"u}neisen parameters of these modes that an increase in volume would result in their hardening. Hence, we can not establish a causal connection with the current set of observations.

\begin{figure}[ht!]
	\centering
	\includegraphics[]{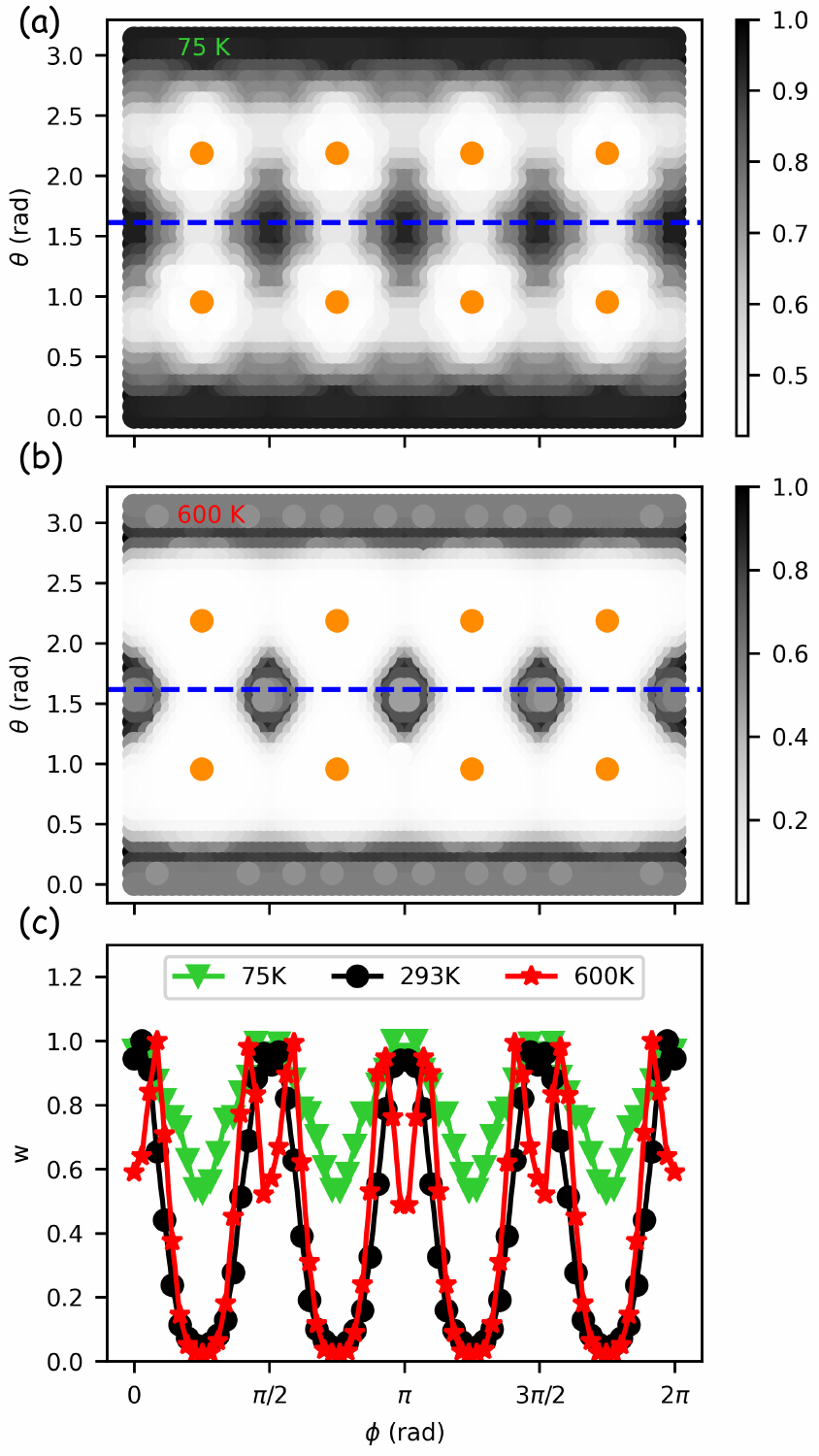}
	\caption{\label{m2p_sphere} Plots showing the $R_{w}$ of different M$_2^+$ OPDs parameterised by spherical coordinates at (a) 75 and (b) 600 K respectively. The greyscale color map corresponds to a value $w = \exp[(R_{min}-R_i)/0.1]$ where $R_{min}$ is the minimum R$_{w}$ for each temperature and the $R_i$ are the $R_{w}$ of each refinement. A value of $w = 1$ then indicates a refinement with a fitting statistic equal to the best fitting statistic across all refinements for that temperature. Thus, darker regions represent better fits than lighter regions. Different color bars are shown for the different temperatures since $R_{min}$ changes with temperature. To aid the eye, points corresponding to an order parameter direction with $Im\bar{3}$ symmetry are shown in orange. (c) A cut through of the above plots, shown in (a) \& (b) with a blue dashed line, showing the variation in the value $w$ for a fixed value of $\theta$. Equivalent plots for the remaining temperature points are shown in the SI.}  
\end{figure}

Next, we consider the sequence of phase transitions in ReO$_3$ under pressure, since these are likely to be indicative of the character of the soft modes observed in the present study. The nature of the initial phase transition with pressure has come under question. The consensus in the literature is that at 5 kbar and 300 K, ReO$_3$ undergoes a transition to a tetragonal $P4/mbm$ structure involving an in-phase octahedral tilt along one pseudo-cubic axis, with only one arm of the propogation vector active (M$_2^+$ $(a;0;0)$). At 5.3 kbar it undergoes a further transition to a cubic $Im\bar{3}$ structure, with all 3 arms of the propogation vector active (M$_2^+$ $(a;a;a)$) and a tilt of equal amplitude along all three pseudo-cubic axes. In Glazer notation, this corresponds to a transition from an $a^+b^0b^0$ tilt system to $a^+a^+a^+$. Some experiments, however, report that there is no transition to the $P4/mbm$ structure\cite{Chatterji2006}, only observing the transition to the $Im\bar{3}$ phase.

Since we expect the distortion which is responsible for the phase transition to be a soft mode at ambient pressure, we can interrogate our PDFs to see if a precursor signature of this phase transition is already present. To do this, we parameterised the three dimensional M$_2^+$ order parameter direction (OPD) $(a;b;c)$ in terms of spherical polar coordinates with $a = r\cos\phi\sin\theta$, $b = r\sin\phi\sin\theta$ and $c = r\cos\theta$. Refinements were performed at fixed values on a grid covering the range of values of $\theta$ and $\phi$, whilst allowing the amplitude of the mode to vary. In this parameterisation, the OPD corresponding to $P4/mbm$ symmetry occurs if both $\theta$ and $\phi$ are integer multiples of $\pi /2$, and for all values of $\phi$ when $\theta = 0,\pi$. The $Im\bar{3}$ OPD would occur when $\phi = \pi/4, 3\pi/4, 5\pi/4$ or $7\pi/4$ and $\theta =\arctan(\pm \sqrt{2})$.

In the Landau theory of phase transitions, the free energy expansion is written as a linear combination of sets of polynomials in the components of the order parameter. These polynomials must be invariant under all of the symmetry operations of the parent space group, $Pm\bar{3}m$ in this case. Since the invariant polynomial truncated at the second (harmonic) order is of the form $a^2 + b^2 + c^2$, which in the spherical coordinate parameterisation is equivalent to $r^2$, the anisotropic $R_w$ distribution we observe over the spherical surface (Fig \ref{m2p_sphere}), is indicative of significant anharmonicity.

At low temperatures, the lowest $R_{w}$ refinements are clustered around points corresponding to the $P4/mbm$ OPD (Fig \ref{m2p_sphere} (a)), while the worst fitting refinements are clustered around points corresponding to $Im\bar{3}$ symmetry. Halfway in between points described by $P4/mbm$ and $Im\bar{3}$ symmetry, corresponding to the OPD $(a;a;0)$ with $I4/mmm$ symmetry, a small improvement in the quality of fit is observed compared to the OPD with $Im\bar{3}$ symmetry. These areas correspond to distortions involving tilts about two orthogonal axes. This likely reflects a saddle point in the energy between the best ($P4/mbm$) and worst ($Im\bar{3}$) fitting OPDs. The anisotropy in the fitting statistics, that becomes more evident at higher temperatures and at large mode amplitudes, points towards significant anharmonicity, since, as discussed above, at the harmonic (quadratic) level, all OPDs must be equivalent with respect to the free energy expansion. Comparing OPDs against our PDF data for $Im\bar{3}$ and $P4/mbm$ phases also shows a clear preference for $P4/mbm$ at all temperatures, with an $R_w$ of 7.28 \% for $P4/mbm$ and 7.83 \%  for $Im\bar{3}$ at 293 K, respectively (Fig S3 in the SI). This supports the consensus that this distortion is the one first reached on the application of moderate pressure. However, it is surprising that even at high temperatures, such a pronounced anisotropic signature of this OPD is evident. We investigate the origin of this anisotropy below.

\begin{figure}[t]
	\centering
	\includegraphics[]{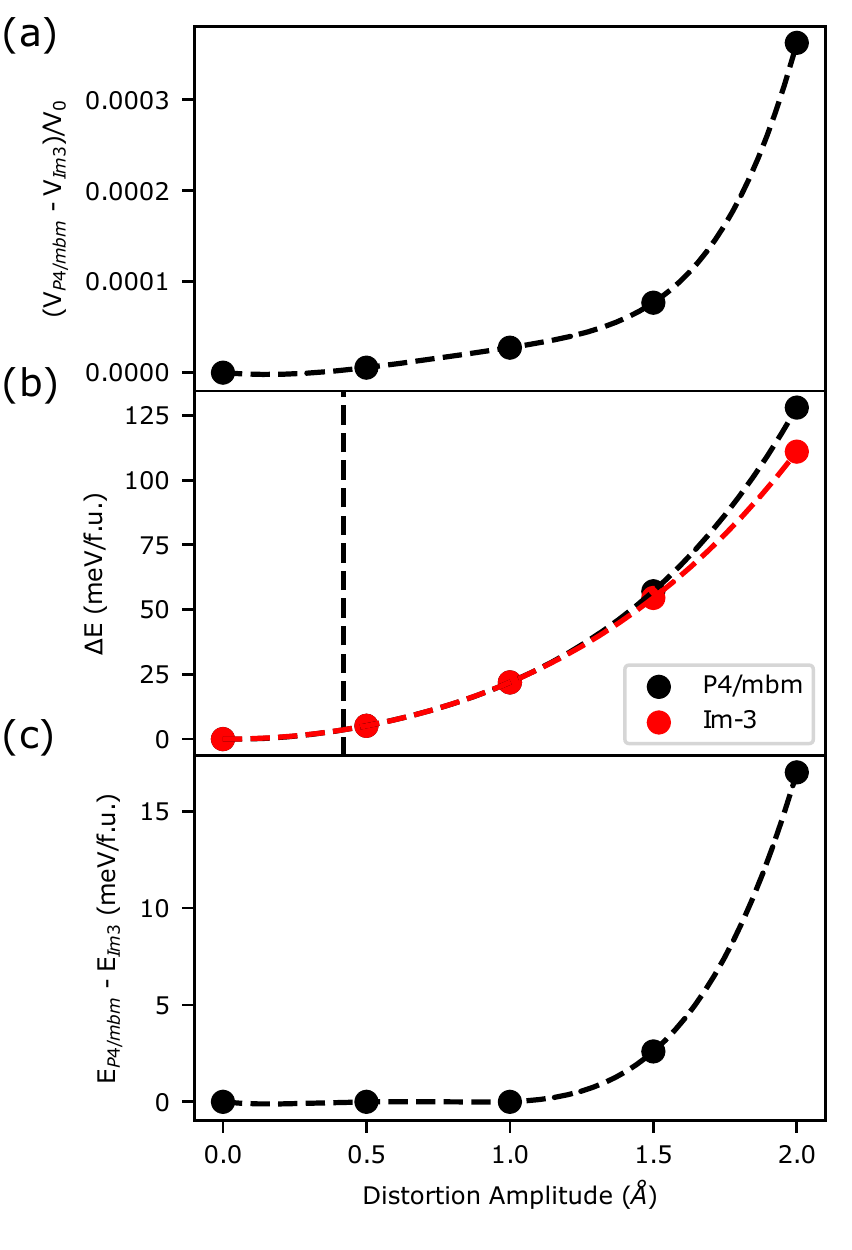}
	\caption{\label{m2pdft}(a) The volume difference (normalised to the volume of the undistorted structure), (b) the energy increase and (c) the energy difference for M$_2^+$ OPDs with $P4/mbm$ and $Im\bar{3}$ symmetries for increasing $2\times 2 \times 2$ supercell-normalised mode amplitudes. These values are obtained using DFT calculations. Lattice parameters were relaxed, respecting the space group symmetry. The vertical line shows the mode amplitude for M$_2^+$ at 293 K (Fig \ref{modeamps})  The energy difference between the two is plotted (c). The lines shown in (b) are 4$^{th}$ order polynomial fits to the data points shown. The line in (c) is the difference between the fits in (b). The line in (a) is shown as a guide to the eye.}
	
\end{figure}
For small mode amplitudes ($<$ $\SI{0.5}{\angstrom}$) of the M$_2^+$ $(a;0;0)$ and $(a;a;a)$ OPDs, such as we observe here, the distortions give rise to an almost identical volume strain in our DFT calculations (Fig \ref{m2pdft} (a)). The two distortions are also found to be equally soft in energy over the range of amplitudes we expect to sample dynamically (Fig \ref{m2pdft} (b\&c)), hence these calculations do not explain the observed anisotropy with respect to the dynamic displacements transforming as M$_2^+$. Possible strain coupling is accounted for in the DFT calculations since the lattice parameters of each distorted structure were relaxed during the energy calculations. The only factor not accounted for by the DFT is that the M$_2^+$ $(a;a;a)$ OPD can couple to displacements that transform as the M$_1^+$ irrep, but this would only serve to decrease the energy of the $Im\bar{3}$ phase further, and so cannot explain why we observe the $P4/mbm$ distortion dynamically at ambient pressure. Since the ground state DFT calculations fail to account for the anisotropy in the OPDs, we must consider these effects to be either due to entropy or anharmonic couplings between phonons. We discount the latter proposal as the anisotropy persists even down to the lowest temperature, where we have shown that any dynamic deviations are small and are accounted for almost exclusively by the M$_2^+$ RUM (Fig \ref{m2ppc}). Hence, in the following paragraph, we explore the prior suggestion that entropy might dictate the anisotropy of the dynamic fluctuations.

By virtue of our experimental observation of the anisotropy of these modes, we have already shown them to be anharmonic in nature. For the OPD M$_2^+$ $(a;0;0)$ with $P4/mbm$ symmetry, the structural fluctuation, taken in the static limit, has additional degrees of freedom that may be realised as dynamic tilts in the directions perpendicular to the spontaneously condensed tilt, corresponding to a line in the phonon dispersion curve between the R and M points in aristotypical perovskite symmetry. The $Im\bar{3}$ structure has tilts along all three pseudo-cubic axes, and consequently will have no additional degrees of freedom of this manner. This leads to the $P4/mbm$ phase, or indeed an anharmonic distortion of this character, having the greater vibrational entropy of the two. This means that it is favoured over other OPDs for the smaller distortion amplitudes that are realised at lower pressures or during dynamic, anharmonic fluctuations of the system. 

The above scenario implies that we should be sensitive in our PDF analysis to low amplitude, harmonic distortions of RUM character that are orthogonal to the dominant anharmonic one. Indeed, we can see a signature of this from the spherical polar coordinate plots of the M$_2^+$ OPD at 600 K, in the form of "rings" around the $(a;0;0)$ OPD (Fig \ref{m2p_sphere} (b)), corresponding to smaller amplitude tilts about the axes orthogonal to the propagation vector of the anharmonic RUM. This is further supported by a fit to the 600 K PDF data with a two-phase model. Each phase contained a large amplitude ($\SI{0.6}{\angstrom}$) M$_2^+$ $(a;0;0)$ distortion, the "anharmonic part", and one smaller amplitude distortion ($<$ $\SI{0.3}{\angstrom}$) of M$_2^+$ $(0;b;c)$ or R$_5^-$ $(0;b;c)$ to mimic deviation in the data due to the line of mainly dispersionless harmonic RUMs running from k = [1/2 1/2 0] to [1/2 1/2 1/2]. All non-zero ratios of in-phase to out-of-phase tilts resulted in a slight improvement to the fit compared to an all in-phase model (Fig S6 in the SI).

The importance of entropy in determining the sequence of soft mode phase transitions in Ruddlesden-Popper perovskites has been highlighted by us recently\cite{Pomiro2020}. In this instance, we discussed how the layering in this structure effectively affords phonon modes with an octahedral tilt character (RUMs about axes perpendicular to the layering axis) a greater vibrational entropy than those with an octahedral rotation character (about the layering axis). The sequence of temperature induced soft mode phase transitions we observed in these compounds were consistent with the idea that the entropic cost of ordering a tilt is higher than that of ordering a rotation. Given that the change in entropy associated with the ordering of a single mode is in principle vanishingly small, it is maybe surprising that the phase transition pathway is dictated in this manner. However, it is the associated renormalisation of phonon modes with RUM and quasi-RUM character, occupying a significant volume in reciprocal space, that provides the non-vanishing contribution to the Gibbs free energy, directing the soft mode transition pathway. Our present results tentatively go beyond these ideas in two respects. Firstly, ReO$_3$ is not a layered perovskite, so there is no distinction between octahedral tilts and rotations. However, it is clear that the same arguments about entropy and phonon renormalisation apply when considering the ordering of tilts about one axis compared to tilts about three. Secondly, our results imply that the anisotropic character of the phonon mode is essentially determined by the entropic cost of the anharmonic fluctuation itself, even without this fluctuation reaching the static limit required to initiate a soft mode phase transition.

In conclusion, a symmetry-motivated analysis of the pair distribution functions of ReO$_3$ has shown that the presence of a rigid unit mode allows this material to exhibit NTE, but a lack of flexibility of the structure limits the magnitude and extent of the NTE behaviour. The rigid unit mode has been shown to be anisotropic, displaying a clear preference for an $(a;0;0)$ order parameter direction, even at elevated temperatures, which is consistent with the $P4/mbm$ space group the structure achieves after its phase transition with pressure. We tentatively suggest that the anisotropy we observe in the tilt direction of the M$_2^+$ RUM is effectively determined by the entropic cost of the fluctuation itself.
	
\section*{Acknowledgements}
T.A.B thanks EPSRC for a PhD studentship through the EPSRC Centre for Doctoral Training in Molecular Analytical Science, grant number EP/L015307/1. M.S.S acknowledges the Royal Society for a University Research Fellowship (UF160265). N.C.B acknowledges computational resources from the Hamilton HPC Service of Durham University and the UK Materials and Molecular Modelling Hub (partially funded by the EPSRC project EP/P020194/1). We are grateful to STFC for the provision of neutron beam time at ISIS, supported under experiment number RB1620329\cite{Dove}.

\bibliography{reo3refs2.bib}	

\end{document}